\documentclass[12pt]{article}

\usepackage[nosort]{cite}
\usepackage{cite}
\usepackage{amsmath}
\usepackage{epsfig}
\usepackage{amsfonts}
\usepackage{amssymb}
\usepackage{multirow}
\usepackage{enumerate}

\usepackage{color}

\usepackage{subfigure}

\usepackage{caption}

\newcommand{\be}{\begin{equation}}
\newcommand{\ee}{\end{equation}}
\newcommand{\bee}{\begin{equation*}}
\newcommand{\eee}{\end{equation*}}
\newcommand{\bea}{\begin{eqnarray}}
\newcommand{\eea}{\end{eqnarray}}
\newcommand{\bean}{\begin{eqnarray*}}
\newcommand{\eean}{\end{eqnarray*}}

\newcommand{\lp}{\left(}
\newcommand{\rp}{\right)}

\begin{document}

\setcounter{page}{0}
\thispagestyle{empty}

\begin{flushright}
CERN-PH-TH/2011-087\\
\today
\end{flushright}

\vskip 8pt

\begin{center}
{\bf \LARGE {
Natural Cold  Baryogenesis from Strongly \\
\vskip 8pt
Interacting  Electroweak Symmetry Breaking \\
 }}
\end{center}

\vskip 12pt

\begin{center}
 {\bf Thomas Konstandin$^{a}$ and G\'eraldine  Servant$^{a,b}$ }
\end{center}

\vskip 20pt

\begin{center}

\centerline{$^{a}${\it CERN Physics Department, Theory Division, CH-1211 
Geneva 23, Switzerland}}
\centerline{$^{b}${\it Institut de Physique Th\'eorique, CEA/Saclay, F-91191 
Gif-sur-Yvette C\'edex, France}}
\vskip .3cm
\centerline{\tt tkonstan@cern.ch, geraldine.servant@cern.ch}
\end{center}

\vskip 13pt

\begin{abstract}
\vskip 3pt
\noindent

The mechanism of ``cold electroweak baryogenesis'' has been so far
unpopular because its proposal has relied on the ad-hoc assumption of
a period of hybrid inflation at the electroweak scale with the Higgs
acting as the waterfall field.  We argue here that cold baryogenesis
can be naturally realized without the need to introduce any slow-roll
potential.  Our point is that composite Higgs models where electroweak
symmetry breaking arises via a strongly first-order phase transition
provide a well-motivated framework for cold baryogenesis. In this
case, reheating proceeds by bubble collisions and we  argue that
this can induce changes in Chern-Simons number, which in the presence of new
sources of CP violation commonly lead to baryogenesis. We illustrate
this mechanism using as a source of CP violation an effective
dimension-six operator which is free from EDM constraints, another
advantage of cold baryogenesis compared to the standard theory of electroweak baryogenesis.
Our results are general as they do not rely on any particular UV
completion but only on a stage of supercooling ended by a first-order
phase transition in the evolution of the universe, which can be natural if
there is nearly conformal dynamics at the TeV scale. Besides,
baryon-number violation originates from the Standard Model only.

\end{abstract}

\newpage

\tableofcontents

\vskip 13pt

\section{Introduction\label{sec:intro}}

There are two major approaches to explain the matter-antimatter
asymmetry of the universe.  One is called {\it leptogenesis}, in which
case the asymmetry is produced by the decay of right-handed neutrinos
in the early universe. The underlying assumptions are that neutrinos
are Majorana particles and that the reheating temperature of the
universe has to be at least $ \sim10^{11}$ GeV.  Leptogenesis is
theoretically well-motivated (for a review see 
e.g.~\cite{Davidson:2008bu}), the main drawback is the difficulty to test
this mechanism, in particular because the CP violation involved in
this process does not manifest itself in low-energy experiments. The
fact that the required reheating temperature of the universe is quite
large may also be a concern, as we will further develop in this paper.

A second avenue, the so-called {\it electroweak baryogenesis}
mechanism \cite{Schmidt:2011zz}, is to consider that the 
matter-antimatter asymmetry has nothing to do with lepton number violation
and is produced at the electroweak (EW) epoch \cite{Kuzmin:1985mm}. In
this case, the sole source of baryon number violation is from the
Standard Model sphalerons. Since sphaleron processes are at thermal
equilibrium before the electroweak phase transition and are
exponentially suppressed in the EW broken phase as 
$\sim \exp\left(
  {-\sqrt{\frac{4 \pi }{\alpha_w}} {\cal C} \frac{\langle \phi(T) \rangle }{T}}
\right)$, where $1.5 \lesssim {\cal C} \lesssim 2.7$ depends on the Higgs mass,
the common lore is that an asymmetry can only be generated
during the EW phase transition, provided that it is first-order.  The
process is non-local as it relies on charge transport in the vicinity
of the CP-violating bubble walls.  Because it involves EW scale
physics only, this mechanism is particularly appealing and will start
to be tested at the LHC. While it is relatively easy to modify the
Higgs potential so that the EW phase transition is strongly
first-order, a main difficulty of EW baryogenesis is that it requires
large new sources of CP violation which are typically at odds with
experimental constraints from electric dipole moments.
 
EW baryogenesis has been investigated in detail mostly in the Standard
Model \cite{Joyce:1994zn} (where it is excluded by now
\cite{Csikor:1998eu, Gavela:1993ts}) and its supersymmetric
extension~\cite{Huet:1995sh, Cline:2000nw, Carena:2002ss,
  Konstandin:2005cd, Cirigliano:2006dg}. The nature of the EW phase
transition has also been studied in models of technicolor
\cite{Cline:2008hr}, although no full calculation of the asymmetry has
been carried out in this context.  The starting observation for this
paper is that in models where the EW symmetry is broken by strong
dynamics, the EW phase transition is typically too strongly
first-order \cite{Creminelli:2001th, Randall:2006py,
  Nardini:2007me,Konstandin:2010cd}, leading to supersonic bubble
growth which suppresses diffusion of CP violating densities in front
of the bubble walls, thus preventing the mechanism of EW baryogenesis
\cite{Espinosa:2010hh}.  Our goal is to revive another mechanism,
known as {\it cold baryogenesis} \cite{Krauss:1999ng,
  GarciaBellido:1999sv, Copeland:2001qw, Cornwall:2001hq, Smit:2002sn,
  GarciaBellido:2003wd, Tranberg:2003gi, vanTent:2004rc,
  vanderMeulen:2005sp, Tranberg:2006dg, Enqvist:2010fd} and show that
it is theoretically well-motivated and only relies on the existence of
a nearly conformal sector at the TeV scale, something which will be
tested at the LHC.  Our conclusions will be  general and
model-independent.  One major advantage of cold baryogenesis is that
it does not depend on the details of the new sources of CP violation,
which can be described by dimension-six effective operators which are
totally unconstrained by EDMs.
 
The cold baryogenesis mechanism is interesting in that it also only
invokes Standard Model baryon number violation and beautifully makes
use of the global texture of the $SU(2)$ electroweak theory.
Nevertheless, so far, it has not received too much acclaim because it
relies on a somewhat unnatural assumption: a period of low-scale (EW
scale) hybrid inflation with the Higgs as the waterfall field.  The
end of inflation is triggered when the Higgs mass turns negative and a
spinodal instability gives rise to an exponential growth of soft Higgs
modes. At this stage, all particles present before low scale inflation
have been inflated away and the universe is cold and empty.
Subsequently, the vacuum energy stored in the Higgs and inflaton
fields reheats the plasma. This energy transfer happens far away from
equilibrium, which makes baryogenesis during this period feasible. One
of the weaknesses of this scenario is that low scale inflation
requires a significant amount of tuning in the inflaton
sector~\cite{Copeland:2001qw,vanTent:2004rc, Enqvist:2010fd}. Besides,
like for the Higgs, a fundamental light scalar inflaton implies a
hierarchy problem.
 
The purpose of the present paper is to demonstrate that the conformal
phase transition in some models of strongly coupled electroweak symmetry
breaking can lead quite generically to a situation in which cold
electroweak baryogenesis is feasible. We want to keep the discussion
as model-independent as possible.  In addition to a nearly conformal potential for the dilaton, 
we only need to assume 
a sizable coupling between the dilaton and the Higgs as well as a slightly 
larger potential energy associated with the dilaton. Let us for instance consider
a  scalar potential of the  type
\be 
\label{eq:potential}
V(\mu,\phi)= \mu^4 \times \left( \
P(({\mu}/{\mu_0})^{\epsilon})+ {{\cal V} (\phi)}/{\mu_0^4} \ \right),
\ee 
where $\mu$ is the  canonical radion (dilaton) field which acquires a vev 
$\mu_0\sim {\cal O}$(1 TeV).  At the confining scale $\mu_{0}$, an
approximate conformal symmetry governs the dynamics.  $|\epsilon| $
parametrizes the explicit breaking of conformal invariance and we are
working in the limit $|\epsilon| \ll 1$ leading to a very shallow
potential $P(({\mu}/{\mu_0})^{\epsilon})$ with widely separate
extrema.   The Randall--Sundrum
model~\cite{Randall:1999ee} with Goldberger--Wise
stabilization~\cite{Goldberger:1999uk} is an explicit realization of
this scenario where the stabilization of a warped extra dimension
solves the hierarchy problem.  It is dual, via the AdS/CFT
correspondence, to a 4D theory where confinement is induced by an
interplay of weakly coupled operators perturbing a
CFT~\cite{ArkaniHamed:2000ds, Rattazzi:2000hs}.  As well-known from
lattice studies, confining phase transitions are first-order for the
rank of the $SU(N)$ gauge group $N\gtrsim 3$ (the exact bound depends
on the matter content) and growing more strongly first-order as $N$
increases. 

For our discussion, we do not need to specify the form of the Higgs
potential $ {{\cal V} (\phi)}$, which can be Standard-Model like.  The
cosmological properties of the potential (\ref{eq:potential}) are
reviewed in a companion article \cite{Konstandin_Servant1}.  The
radion acts in this context similar to an inflaton and the conformal
symmetry protects the Higgs as well as the radion mass thus solving
the hierarchy problem.  For example, let us consider the
Randall-Sundrum scenario with the 5D warped metric 
$ds^2=e^{-2r/l} \eta_{\rho\sigma}dx^{\rho}dx^{\sigma}+dr^2$ (the radion field is then defined as $\mu=l^{-1}e^{-r/l}$ where $l\sim M_{Pl}^{-1}$ is the AdS$_5$ curvature) and with the 
 Standard Model  Higgs field $\varphi$ on the
infrared brane. Because the radion arises as a gravitational degree of freedom, its coupling to the Higgs 
 arises from the induced metric on the IR brane.
The resulting 4D effective action for the Higgs is then:
\be
{\cal L}_4= \frac{\mu^2}{\mu_0^2} g^{\rho\sigma}D_{\rho}\phi D_{\sigma} \phi -\frac{\mu^4}{\mu_0^4} \frac{\lambda}{4}(\phi^2-v_0^2)^2
\label{HiggsonIRbrane}
\ee
where the Higgs $\varphi$ has been redefined as $\phi= \mu_0 l \varphi$ and
 $v_0= \mu_0 l v$ where $v\sim M_{Pl}$.  We recover the second term in the potential (\ref{eq:potential}) although for a non-canonical Higgs field $\phi$.
In the limit $\langle \mu\rangle \rightarrow 0$, one can easily see that the vev of the canonical Higgs is proportional to $\mu$ and therefore  $\langle \mu\rangle $
sets the EW scale. 
We also note that in the case of a little hierarchy $\mu_0 >v_0$, we can neglect the Higgs contribution when studying the dynamics of the radion.
In general (for instance if the Higgs is delocalized towards the bulk), the interactions between the Higgs and the radion will be different from (\ref{HiggsonIRbrane}) but this should not affect much our discussion.
What we rely on in this paper is that
the conformal phase
transition leading to $\langle \mu \rangle \neq 0$ is strongly first-order and proceeds by bubble
nucleation. This modifies significantly the standard picture of
reheating.

In the next section, we review the microscopic picture of cold
electroweak baryogenesis.  In Section~\ref{sec:preheat} we discuss
preheating after a stage of supercooling ended by a strongly first-order
phase transition and argue that models with nearly conformal dynamics
offer all the required conditions for successful cold baryogenesis. We
estimate the resulting baryon asymmetry in Section~\ref{sec:baryo} and
conclude in Section~\ref{sec:dis}.

\section{Cold electroweak baryogenesis\label{sec:CEWBG}}

The main idea of cold baryogenesis relies on the evolution of winding
number and Chern-Simons number in a fast tachyonic EW
transition.  In the `standard' picture (see
e.g.~\cite{GarciaBellido:1999sv}), the EW phase transition is
triggered by a rapid change in the Higgs mass (``quenching") in a
nearly empty Universe. This can be arranged for instance in a
low-scale inverted hybrid inflation scenario where the inflaton is
coupled to the Higgs~\cite{Felder:2000hj, GarciaBellido:2002aj,
Smit:2002sn, GarciaBellido:2003wd, Tranberg:2003gi}. The resulting
tachyonic instability leads to strongly out-of-equilibrium conditions
with an exponential growth of occupation numbers in the Higgs fields
and after a short while the system becomes classical. The $SU(2)$
orientation of the Higgs field is inhomogeneous in space such that
different regions approach different minima in the Higgs potential,
similar to a spinodal decomposition.  The dynamics of the system can
lead to substantial changes in the Chern-Simons number of the $SU(2)$
gauge fields
\be
N_{CS} = -\frac{1}{16 \pi^2} \int d^3x \,
\epsilon^{ijk} \, \textrm{ Tr } \left[
A_i \left( F_{jk} + \frac{2i}{3} A_j A_k \right) \right],
\ee
and can therefore induce baryon number violation via the quantum
anomaly that relates a change in baryon number $B$ to a change in
Chern-Simons number $N_{CS}$
\be
\Delta B= 3 \Delta N_{CS}.
\ee
The key point is that the dynamics of the Chern-Simons number is
linked to the dynamics of the Higgs field via the Higgs winding number
\be
N_H=\frac{1}{24 \pi^2}
\int d^3 x \, \epsilon^{ijk} \,
\textrm{ Tr } \left[
\partial_i \Omega \Omega^{-1}
\partial_j \Omega \Omega^{-1}
\partial_k \Omega \Omega^{-1} \right],
\ee
where $\Omega $ is given by the elements of the usual $SU(2)$ Higgs
doublet $\phi$ of the SM :
\be
\frac{\rho}{\sqrt2} \, \Omega  = ( \epsilon \phi^*, \phi) =
\begin{pmatrix}
\phi_2^* & \phi_1 \\
-\phi_1^* & \phi_2 \\ 
\end{pmatrix} \ \  , \ \  \rho^2=2 (\phi^*_1 \phi_1 + \phi^*_2 \phi_2).
\ee
Both the winding number and the Chern-Simons number change under large
gauge transformations. However, the variations $\Delta N_{CS}$,
$\Delta N_H$ and the difference 
\be
\delta N \equiv N_{CS} - N_H,
\ee
are gauge invariant.  In the vacuum, $\delta N =0$. A texture is a
configuration which has $\delta N \not= 0$, with a Higgs length $\rho$
that is equal to its vacuum value everywhere and which only carries
gradient energy. In the absence of gauge fields, textures are not
stable configurations but shrink quickly~\cite{Turok:1989ai} and the
vacuum configuration is the constant Higgs field with vanishing
winding number.

Cold EW baryogenesis is based on gauged textures of the
EW gauge sector of the SM~\cite{Turok:1990in}.  A gauged
texture is also unstable and its evolution depends on its length scale
$L$. A localized texture just collapses by changing $N_H$, in which
case baryon number is not violated. However, if textures are spread
out and larger than the size of gauge fields $\sim m_W^{-1}$, gauge
fields can cancel the Higgs gradient energy and textures decay by
changing the Chern-Simons number (thus producing baryon
number)~\cite{Lue:1996pr}. For example, consider the configuration
\be
\label{eq:H_wind_ex}
\Omega (x_\mu) =  \frac{ \, \mathbf{1}L +   \sigma_i (x_i - x^0_i)}
{\sqrt{L^2 + (x-x^0)^2}},
\ee
which carries a non-trivial winding number ($\sigma_i$ are the Pauli
matrices, $L$ parametrizes the size of the configuration and $x_i$ its
position). In order to produce such a configuration out of a trivial
one ($\Omega=\mathbf{1}$), the Higgs field has to surpass a potential
barrier. While the winding number of the configuration
(\ref{eq:H_wind_ex}) is rather homogeneously spread in space, roughly
half of the winding is localized near the position of the Higgs zero
 for a configuration that is close to surpassing the
barrier~\cite{vanderMeulen:2005sp}. 
This `half-knot' changes sign when the barrier is surpassed
in a way that changes the total Higgs winding by one unit. Since the
system has to approach the vacuum at later stages, the Higgs winding
has to either decay ($L \to \infty$) or be dressed by the gauge fields
($A_i \to \partial_i \Omega \Omega^{-1}$). In the latter case, this leads to a change in
Chern Simons number, hence a change in baryon number.

\begin{figure}[t!]
\centering
\includegraphics[scale=.525]{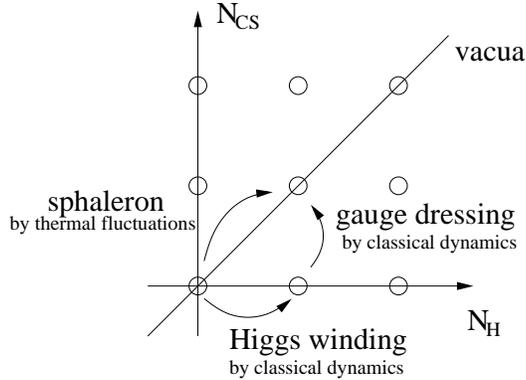}
\caption{
\label{fig:Bviolation}
\small In standard EW baryogenesis, baryon number violation occurs via high temperature induced sphaleron transitions, along the ``vacua" line $N_H = N_{CS}$. In contrast, in cold baryogenesis, sphaleron transitions are switched off and baryon number violation takes place in a two-step process via the production and decay of textures (configurations having $N_H \neq N_{CS}$). First, large kinetic energy stored in the scalar sector induces Higgs winding transitions. Second, these winding configurations can decay by changing Chern-Simons number, thus producing baryon number.}
\end{figure}

The inflaton dynamics can lead to a parametric resonance (preheating)
when large amplitude non-thermal excitations in both the inflaton and
coupled Higgs field arise. During the EW phase transition and the
following preheating stage, $\delta N \neq 0$ configurations are then
abundantly produced.  They eventually relax to zero. However, in the
presence of CP violation, $\delta N > 0$ and $\delta N < 0$ winding
configurations which have a size comparable to $m_W^{-1}$ evolve
differently towards the vacuum.  $\delta N < 0$ windings have a
slight preference to relax by changing $N_H$ while $\delta N > 0$
configurations have a slight preference to unwind by changing
$N_{CS}$. The imbalance between a change in winding number and a change
in Chern-Simons can then generate a net baryon number under out of
equilibrium conditions.  

A common source of CP violation employed in this context is the higher
dimensional operator\footnote{Operators obtained by
integrating out the SM fermions \cite{Shaposhnikov:1986jp,
Shaposhnikov:1987tw, Smit:2004kh, Hernandez:2008db} have also been
advocated as efficient CP violating sources for cold EW
baryogenesis~\cite{Tranberg:2009de, Tranberg:2010af}.  However, the
validity of the approach can been questioned due to the role of hard
modes in the generation of winding number (i.e.~harder than the charm
or strange quark mass which is the inverse of the expansion parameter
in~\cite{Hernandez:2008db}). Furthermore, the operators of
\cite{Hernandez:2008db} could not be reproduced using different
techniques~\cite{GarciaRecio:2009zp, Salcedo:2011hy}. }
\be
\label{eq:CPVop}
{\cal O}_{CPV} =  \frac{1}{M^2} \phi^\dagger \phi \tilde F F,
\ee
which acts as a chemical potential for Chern-Simons number and yields
the required bias towards baryon number generation.  A major advantage
of an operator of the form (\ref{eq:CPVop}) is that the observed
baryon asymmetry can be explained without conflicting with constraints
from electric dipole moments.  

Because of the non-perturbative nature
of the phenomenon, it is difficult to derive reliable analytical
estimates for the baryon asymmetry. However, a nice feature of cold
EW baryogenesis is that most of the process can be simulated
on a computer lattice (from the very early to the very late stages)
~\cite{Rajantie:2000nj, Felder:2000hj, GarciaBellido:2002aj, Smit:2002sn,
GarciaBellido:2003wd, Tranberg:2003gi}.  In particular, the behavior
of winding and Chern-Simons number can be explicitly observed
\cite{vanderMeulen:2005sp}.

A crucial ingredient for a successful baryogenesis mechanism is to
prevent washout of the baryon asymmetry which is possible if the
tachyonic transition takes place in a cold universe. This is generally
achieved by engineering a low scale inflaton coupled to the Higgs.
Our goal in this paper is to show that there is another natural and
well-motivated route for implementing cold baryogenesis: a nearly
conformal phase transition at the TeV scale.

To conclude this section, we point out that an earlier proposal for
{\it local} EW baryogenesis (in which $B$ and $CP$ occur
together in space and time) based on the decay of textures was made in
Ref.~\cite{Turok:1990in, Dine:1990fj, Dine:1991ck, Lue:1996pr}. It
assumed a first-order EW phase transition, proceeding through the
nucleation and expansion of bubbles filled with the new (broken)
phase.  The idea was based on the production and decay of
configurations carrying non-vanishing winding number while the passing
bubble wall drives the system out-of-equilibrium. It was however later
shown that this mechanism does not lead to a sufficient baryon
asymmetry ~\cite{Lue:1996pr}. This problem was cured in the so-called
{\it non-local} (by charge transport) and now standard EW
baryogenesis mechanism~\cite{Cohen:1994ss} where CP violation
originates in the fermionic sector and is transported by diffusion
into the symmetric phase where sphaleron transitions are unsuppressed
and biased by the CP-violating fermion densities, thus producing the
baryon asymmetry.  However, in this case, the asymmetry is produced at
high temperature and leads to strong constraints on the finite
temperature Higgs potential nature via the sphaleron bound $\phi/T
\gtrsim 1$ where $\phi$ and $T$ are evaluated at the nucleation
temperature.  In contrast, in cold baryogenesis, the EW phase
transition takes place at very low temperature (and does not have to
be first-order). 
 $\Delta B$ generation in cold and standard cases are sketched in Fig.~\ref{fig:Bviolation}.

\section{Preheating after a relativistic first-order phase transition\label{sec:preheat}}


For cold electroweak baryogenesis to be viable, one has to ensure that
in the first stage of the reheating process, most of the energy
resides in the scalar sector with momenta of the order of the
electroweak scale before the system approaches equilibrium. This is
the topic of this section.
We have seen earlier that at the EW symmetry breaking transition, the
Higgs field produces winding as it falls towards the vacuum. Ignoring
gauge fields, winding configurations collapse and unwind. In a gauge
theory, the behavior of textures is more complex. The tendency of the
Higgs field to unwind is competed by the gauge fields which can cancel
gradient energy in the Higgs field. In particular, if the initial size
$L$ of the winding configuration is $L< m_W^{-1}$ the gradient term in
the Higgs field is large and pulls the Higgs field over the potential
barrier and changes $N_H$ to match $N_{CS}$ whereas if $L>m_W^{-1}$
the gauge field changes winding number to match $N_H$.  To produce
baryon number, we need to change the value of $N_{CS}$ so what is
crucial is to determine under which conditions winding configurations
can be produced during a first-order phase transition and what is
their typical size.  The first-order phase transition we have in mind
refers to the conformal phase transition described in detail in
\cite{Konstandin_Servant1} which implies that the induced electroweak
symmetry breaking can take place at a temperature below the sphaleron
freeze-out temperature. Since the radion vev induces a Higgs vev, the
bubbles should be understood as both radion and Higgs bubbles.

Before percolation, the energy budget of a first-order phase
transition depends on several factors such as the amount of latent
heat released or the friction in the bubble
wall~\cite{Espinosa:2010hh}. Part of the energy is transformed into
bulk motion and heat in front of the wall or is absorbed by the
particles that climb the potential well induced by the changing scalar
field vev. These are the microscopic processes that counteract the
expansion of the bubble walls and constitute the friction felt by the
scalar field. However, in the case of interest, the universe is cold
and almost empty when the phase transition occurs such that the
surrounding plasma cannot efficiently hinder the wall expansion and
most of the energy is used to accelerate the
wall~\cite{Bodeker:2009qy}, quite similarly to the situation in
vacuum~\cite{Kosowsky:1992vn}. After some time of expansion, the walls
are highly-relativistic  and since their size at the end of the
phase transition is of order of the Hubble constant they can reach
velocities of order $\gamma_w \sim (m_{Pl}/m_W)^{1/2}\sim 10^8 $.

At the end of the phase transition, most of the energy of the system
is localized in the expanding bubbles.  Ultimately, when the bubbles
start to collide and percolate this energy has to reheat the universe
in a feasible cosmological scenario in order to reproduce the
predictions of big bang nucleosynthesis. There are competing effects
that are relevant in this era of equilibration, as e.g.~particle
production~\cite{Watkins:1991zt} or production of classical scalar
waves~\cite{Hawking:1982ga, Watkins:1991zt, Kolb:1996jr,Kolb:1997mz} by reflection
of the coherent bubble walls.

Let us consider the production rate of a fermionic species
that interacts with the scalar field $\phi$ via a Yukawa interaction during
bubble collisions
\be
{\cal L} \ni \, g \, \bar \psi \, \phi \, \psi. 
\ee
The number of particles produced per area is~\cite{Watkins:1991zt}
\be
\label{eq:pprod}
\frac{N}{A} = 2 \int \frac{dk\, d\omega}{(2 \pi)^2} 
\left| \phi(\omega, k)\right|^2 \, {\rm Im }\Gamma^{(2)} (\omega^2 -k^2),
\ee
where $\phi(\omega, k)$ denotes the Fourier transform of the Higgs
field configuration and $\Gamma^{(2)}$ denotes the second derivative
of the effective action that in perturbation theory to leading order
is given by
\be
{\rm Im }\Gamma^{(2)}(p^2) = \frac{g^2 p^2}{8 \pi} 
\left( 1 - \frac{4 m^2}{p^2}\right)^{3/2}
\,\Theta(p^2 - 4 m^2),
\ee
for a fermion $\psi$ of mass $m$. 
We use in the following this expression to estimate the energy 
that is transfered into the fermionic sector during reheating.

The production of sufficient Higgs winding number requires sizable
kinetic energy in the Higgs field in the form of classical
configurations that can surpass the potential barrier. This is
prohibited if the energy of the scalar sector is drained too fast into
fermions. How efficient scalar wave and fermion production is depends
crucially on the vacuum structure of the scalar sector and in the
following we will discuss three relevant cases in the limit of highly
relativistic bubble wall velocities, $\gamma_w \to \infty$, namely
\begin{enumerate}[(a)]
\item a periodic potential
\item a potential with two nearly degenerate minima
\item a potential with two asymmetric minima 
\end{enumerate}
 During the phase transition the dynamics of both fields,
Higgs and radion, are in principle important. The Higgs is most
important for the generation of the baryon asymmetry via Higgs winding
and Chern-Simons number. On the other hand, the radion will dominate
the dynamics of the phase transition due to a slightly larger energy
scale.  Hence, whether the system goes back to the symmetric phase 
after the collision depends mostly on the features of the radion potential.
In the following, we identify the scalar potential under consideration
with the nearly conformal radion potential that belongs to the last
category above.

\begin{figure}[t!]
\centering
\subfigure[(a)]{
\includegraphics[scale=0.425]{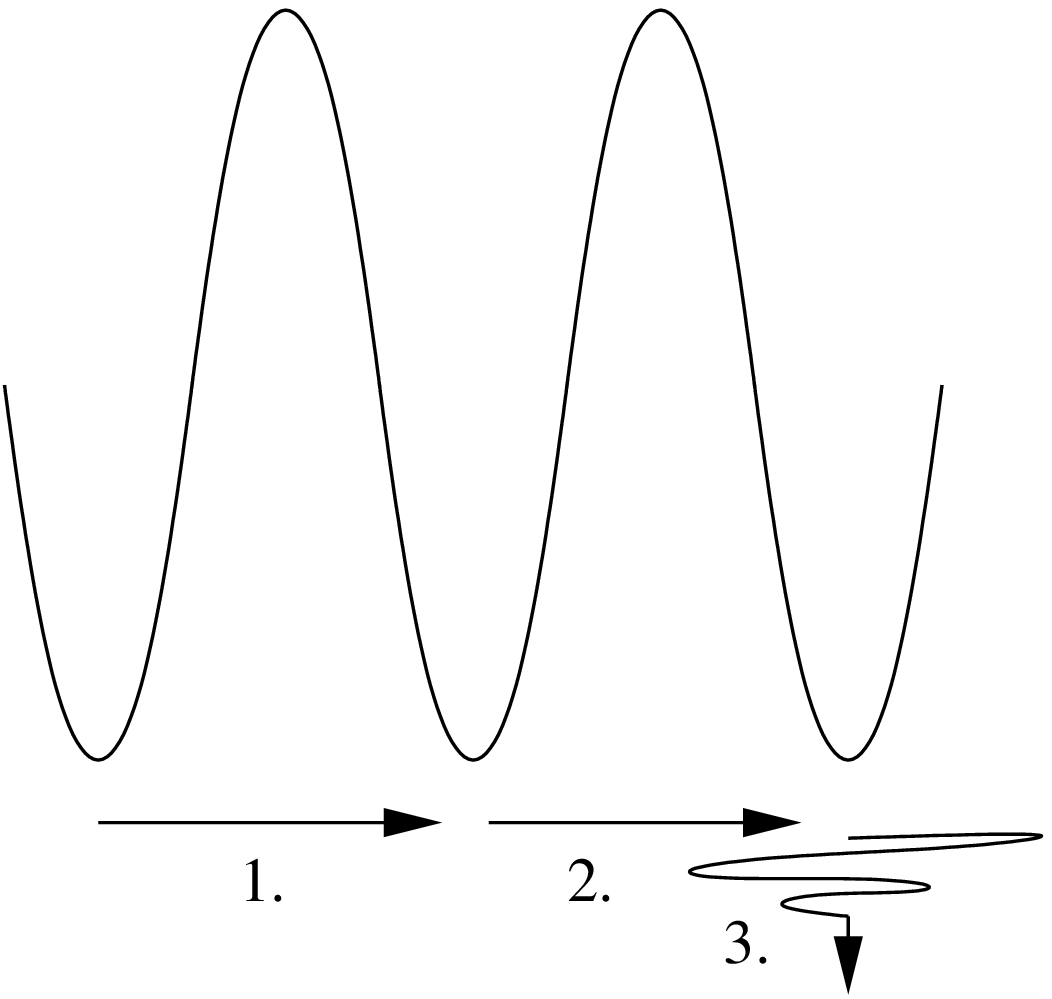}}
\hspace{1cm}
\subfigure[(b)]{
\includegraphics[scale=.425]{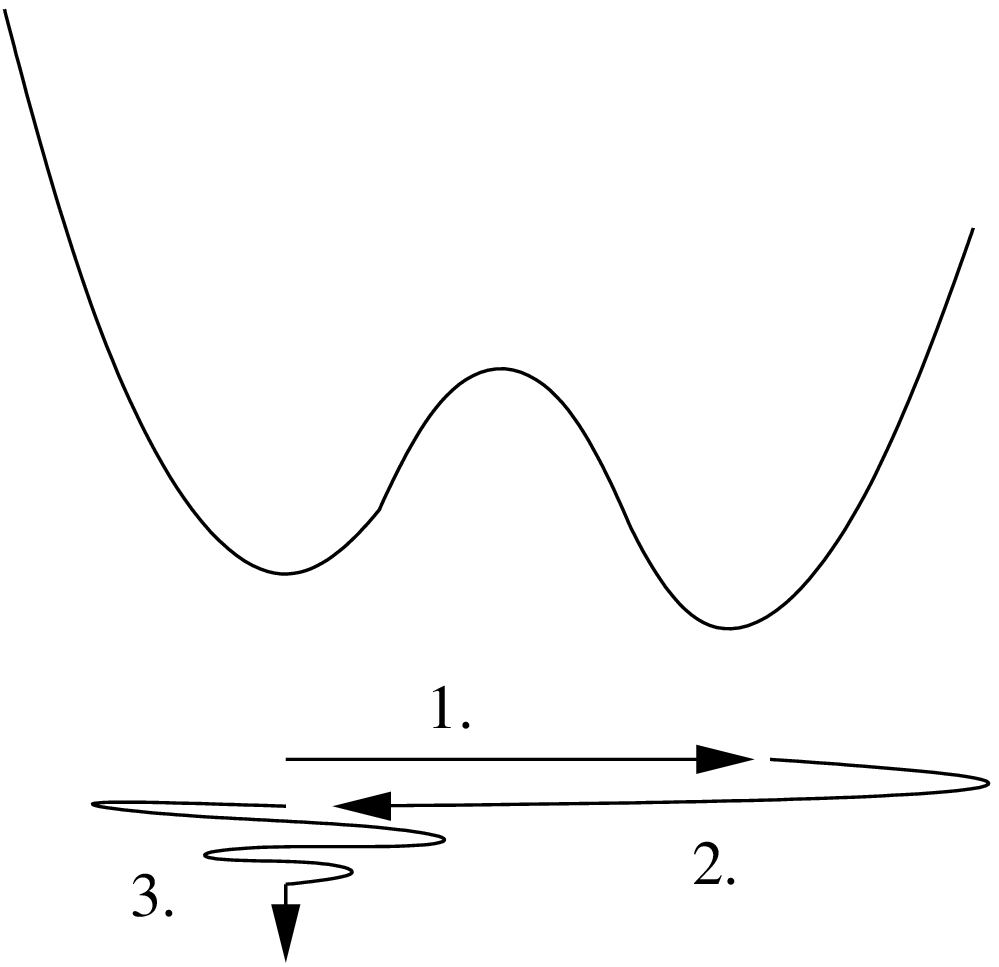}}
\hspace{1cm}
\subfigure[(c)]{
\includegraphics[scale=0.425]{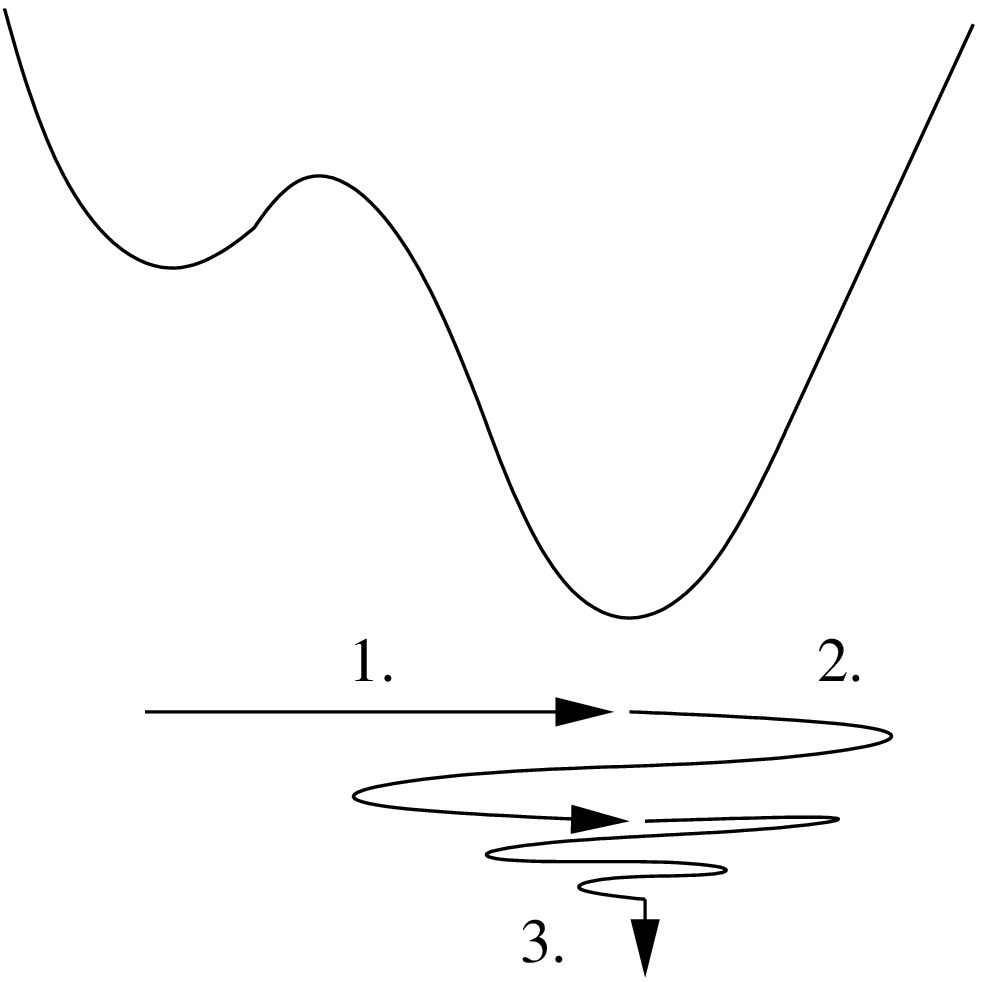}}
\caption{
\label{fig:pots}
\small The path of the scalar field for the three different potentials
a), b), c) discussed in the text.  ``1" denotes the path in the
expanding bubble walls. ``2" is the path during the collision. ``3" is
the path in the collided region.}
\end{figure}

First, if the vacuum structure is periodic, colliding bubble walls do
not reflect and pass freely each other without interfering much (see
Fig.~\ref{fig:pots}(a)). Energy is very slowly carried over from walls
to the scalar (radion and Higgs) sector. Most of the energy decays
into SM particles before it is accumulated in the scalar sector.
Besides, particle production is suppressed by the Lorenz factor
$\gamma_w^2$ of the colliding bubble walls~\cite{Watkins:1991zt}.

\begin{figure}[!t]
\begin{center}
\includegraphics[width=0.75\textwidth, clip ]{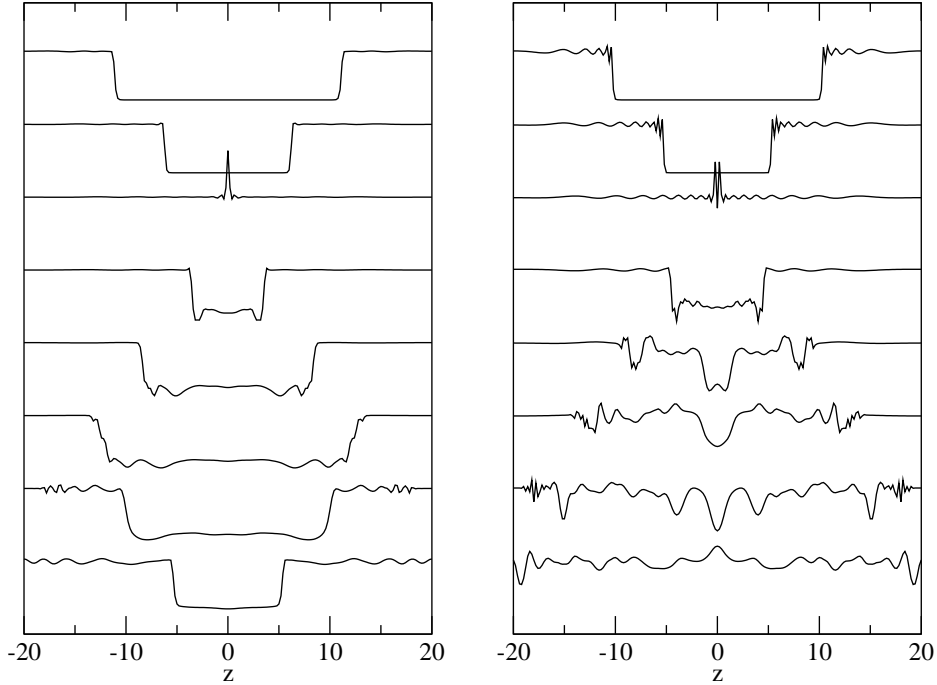}
\caption{
\label{fig:coll_slice} 
\small Collision of planar bubble walls (with initial velocity
$v_w=0.5$) using the potential (\ref{eq:toy_pot}) for
$\lambda=1$. The left (right) plots are respectively for the nearly
symmetric ($\eta=0.2$) and asymmetric ($\eta=0.6$) potentials.  In
case (b), the walls are reflected, and eventually stop expanding until
the symmetric phase collapses again. In case (c) the field cannot
leave the basin of attraction of the broken phase.  These plots
correspond to different slices of the collisions shown in
Fig.~\ref{fig:coll}.}
\end{center}
\end{figure}

\begin{figure}[!ht]
\begin{center}
\includegraphics[width=0.4\textwidth, clip ]{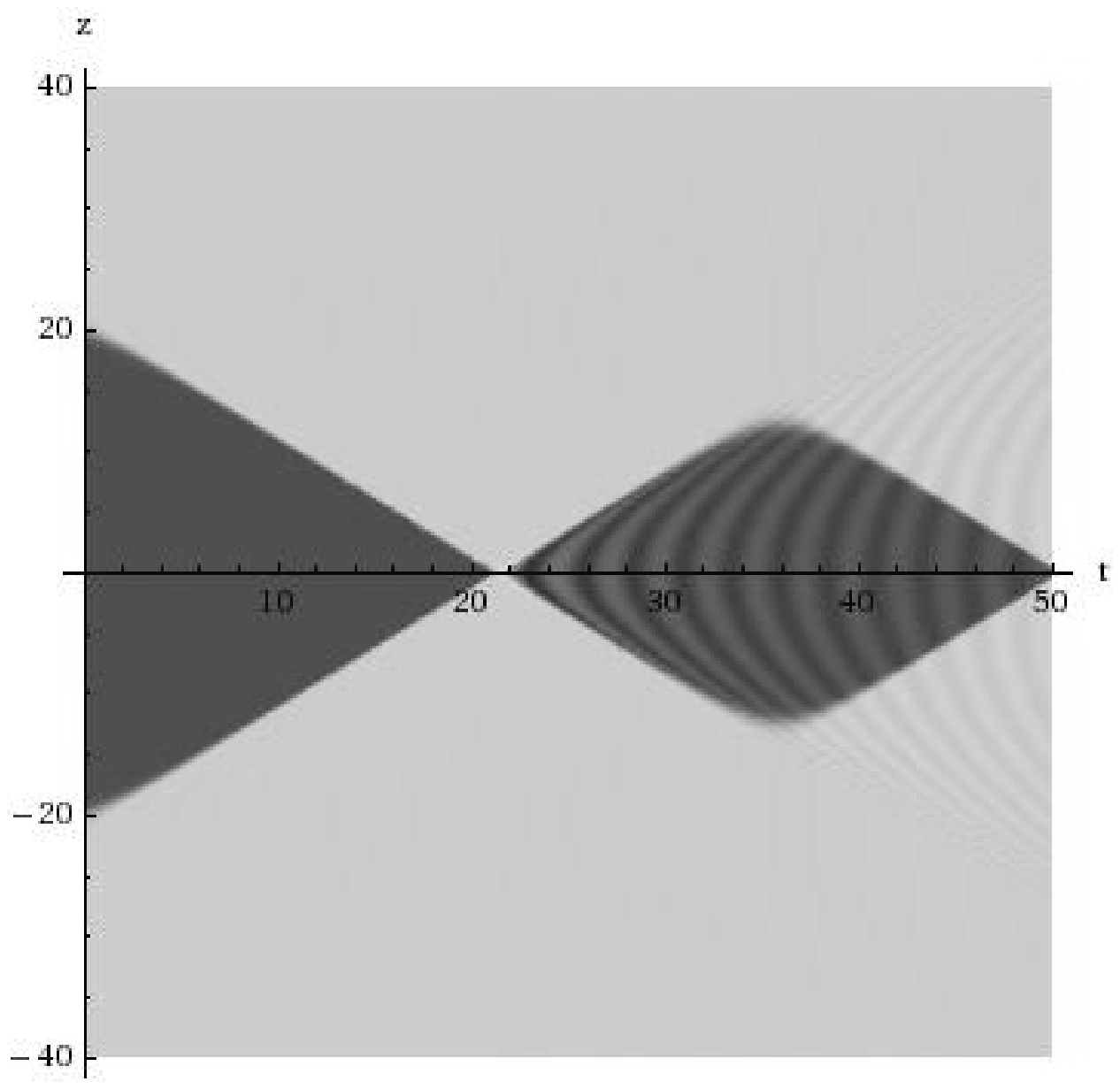}
\includegraphics[width=0.4\textwidth, clip ]{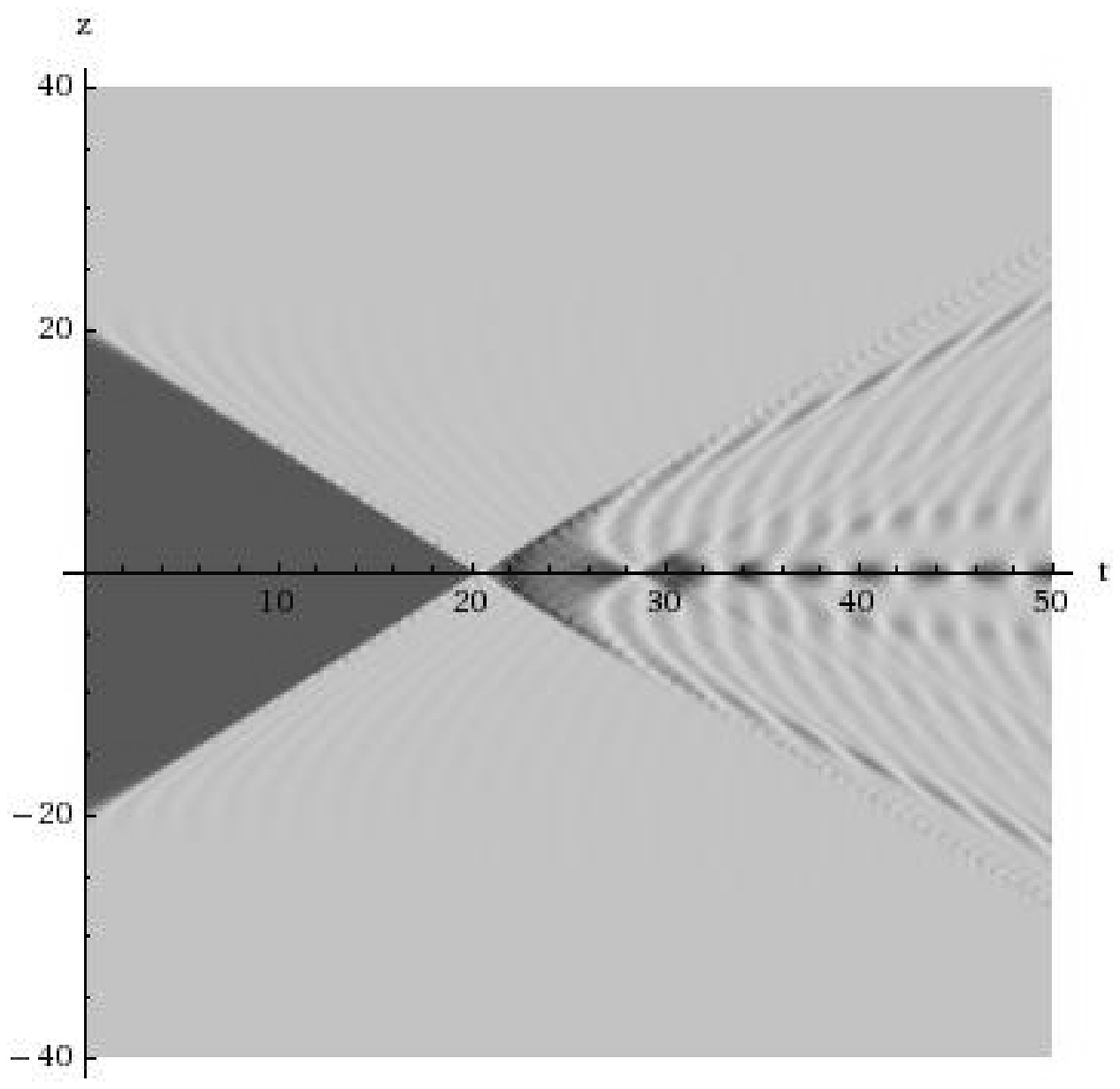}
\includegraphics[width=0.4\textwidth, clip ]{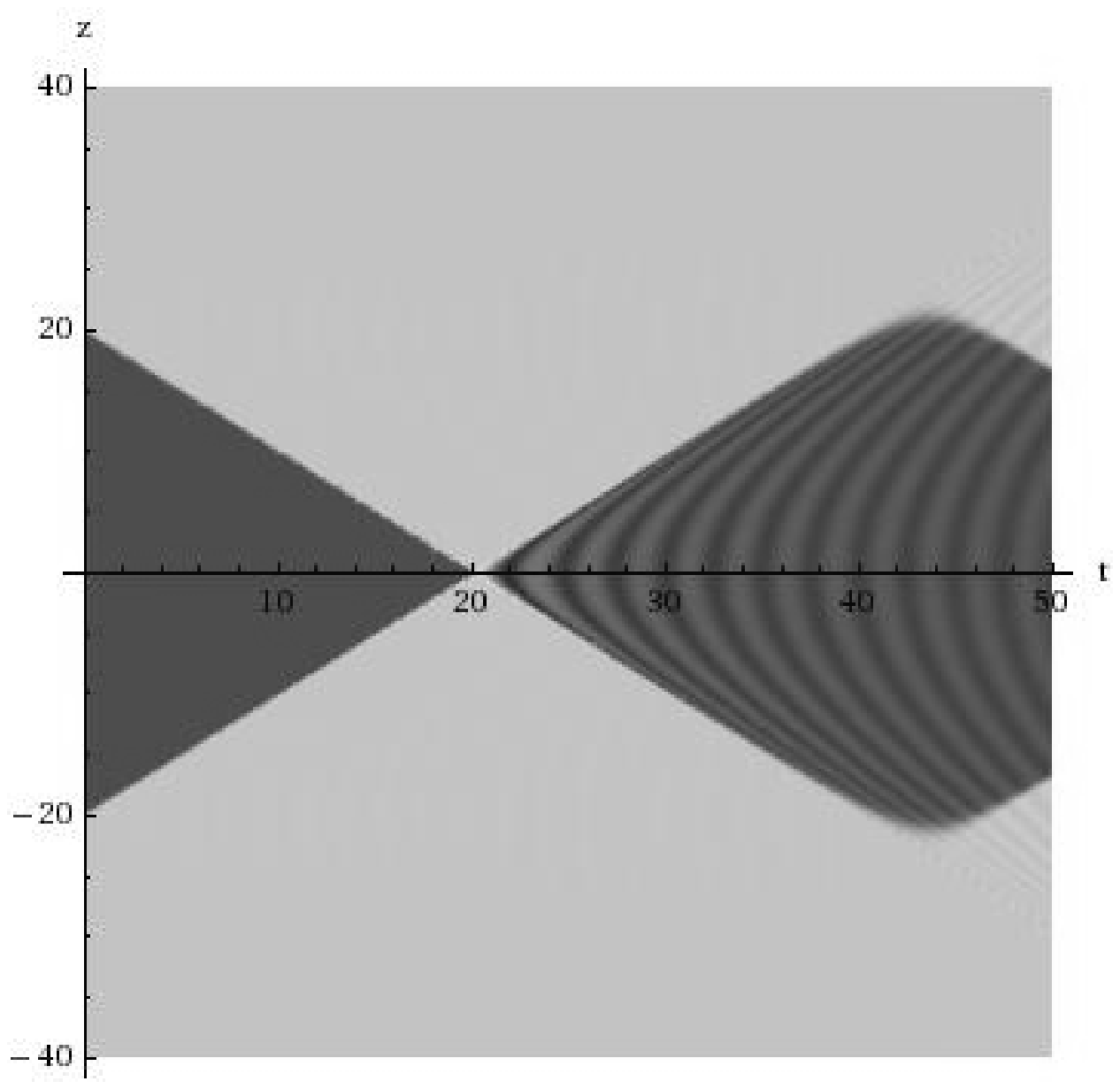}
\includegraphics[width=0.4\textwidth, clip ]{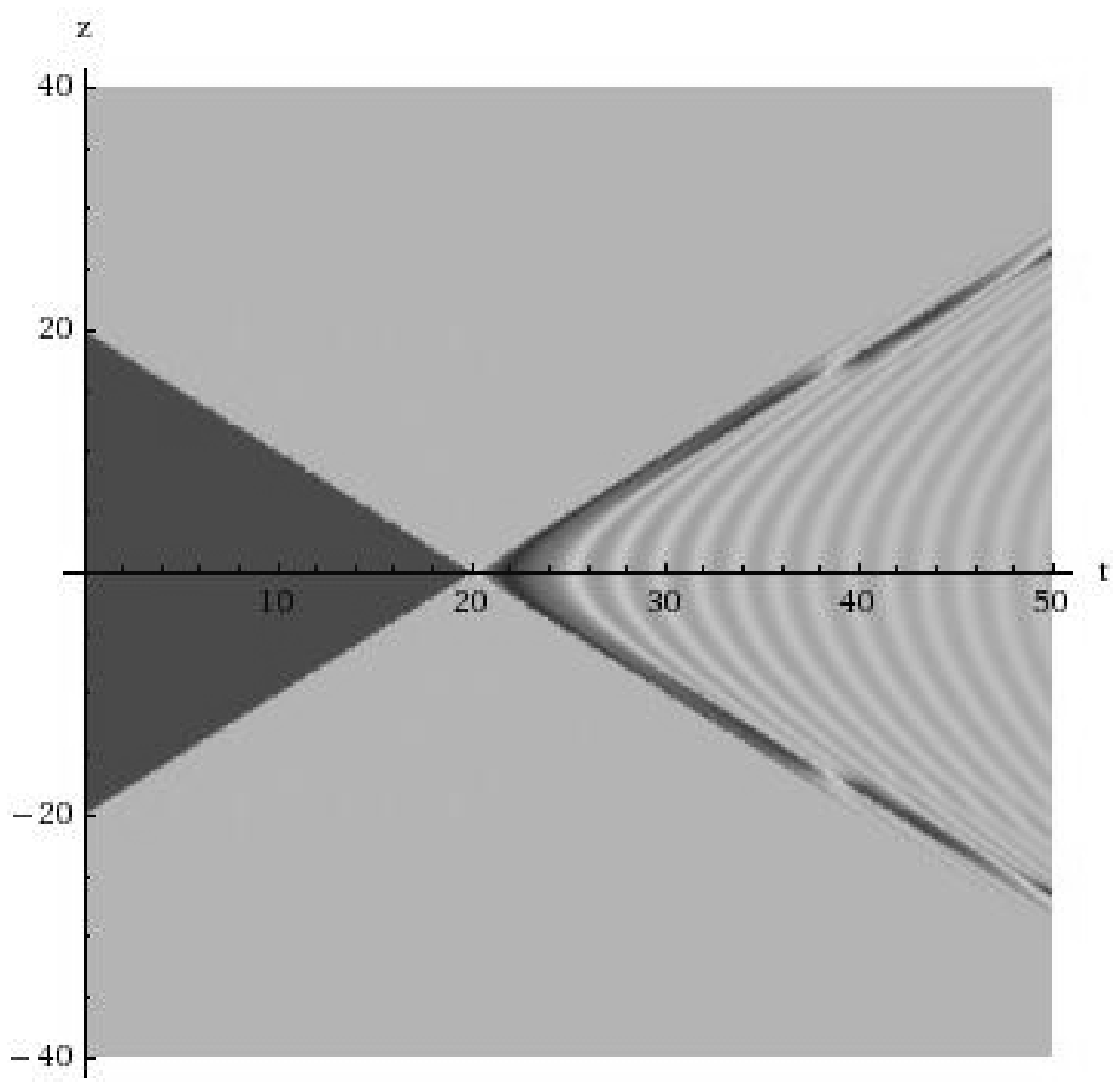}
\includegraphics[width=0.8\textwidth, clip ]{figs2/energies.eps}
\vskip -0.2 cm
\caption{
\label{fig:coll} 
\small Collision of planar bubble walls for the potential
(\ref{eq:toy_pot}) with $\lambda=1$. The top (bottom) plots use as
initial wall velocity $v_w=0.5\, (0.98)$, respectively.  The left
(right) plots are for the symmetric (asymmetric) potential with
$\eta=0.2 \, (0.6)$.  Light (dark) gray corresponds to the broken
(symmetric) phase.  In the left case, the walls are reflected, and
eventually stop expanding until the symmetric phase collapses
again. In the right case, the field cannot leave the basin of
attraction of the broken phase. The last pair of plots shows the time
evolution of the fractions of the total energy in potential energy,
bubble wall energy and ``kinetic" energy of the classical scalar
field in the case $v_w=0.5$ (see text).}
\end{center}
\end{figure}

Secondly, in the case (b) where the scalar potential has two nearly
degenerate local minima, the expanding bubble walls bounce in the
potential and reflect at each other (see Fig.~\ref{fig:pots}(b)). This
reestablishes a region of symmetric phase between the collided bubble
walls. The expansion of the bubble walls is counteracted by the
pressure difference, such that the bubble walls are slowed down and
finally the symmetric phase collapses again (as shown in
Ref.~\cite{Watkins:1991zt} and in the left plots of
Fig.~\ref{fig:coll_slice} and Fig.~\ref{fig:coll}). Each collision
releases some fraction of the wall energy into scalar waves.  Most of
the energy is radiated away after a few collisions.  Even though
expanding bubble walls do not decay into fermions\footnote{This can be
seen by noting that the wall profile has no time-dependence in the
co-moving frame and only a support for $p^2 \leq 0$ in Fourier
space. Hence there is no particle production according to
(\ref{eq:pprod}).}, thermalization occurs by production of scalar
waves. The different collisions are separated by a time of the order
of the Hubble time, which is much longer than the electroweak time
scale that determines the decay rate of the classical scalar
waves\footnote{Using (\ref{eq:pprod}) the decay rate of the classical
Higgs waves is basically the one of the Higgs particle.}. This
constitutes a serious problem for us since the process of transferring
the bubble wall energy into EW scale scalar configurations is very
inefficient.
%
%
On top of that, the reflections of bubble walls themselves lead to
significant particle production: a fixed fraction $g^2$ of the energy
of the colliding walls goes into production of
fermions~\cite{Watkins:1991zt}, even in the limit $\gamma_w \to
\infty$. Hence, in the case of nearly degenerate vacua, a sizable
fraction of the energy will be drained into the fermionic
sector. Therefore, it is questionable that a sizable energy fraction
is present in the form of classical kinetic energy of the Higgs field.

The potential (c) with two asymmetric minima gives different results.
When two scalar bubbles collide, the scalar field bounces and is
reflected close to the symmetric phase. However, a partial loss in
energy implies that the field only approaches the old minimum to a
certain extent. In Ref.~\cite{Giblin:2010bd, Easther:2009ft}, it is
shown that the walls are reflected only if the field can reach the
basin of attraction of the symmetric minimum. If not, the field
bounces back close to the symmetric minimum but remains in the basin
of attraction of the broken phase. In this case, the field approaches
after a short while the broken minimum and starts oscillating around
it (see Fig.~\ref{fig:pots}(c)).  This very much resembles the
situation of the tachyonic instability after low-scale inflation.

There are two relevant parameters that decide in which basin of
attraction the field ends up. For highly relativistic walls, $\gamma_w
\to \infty$, the field tends to bounce closer to the symmetric
phase. The second relevant quantity is the size of the basin of
attraction of the symmetric phase. We checked that for a very shallow
potential with widely separated minima $\mu_+$ and $\mu_- $
\cite{Konstandin_Servant1}, reflection is very unlikely due to the
huge hierarchy $\mu_+/\mu_- \ll 1$ and as a result the basin of
attraction of the symmetric minimum is unattainable.

\begin{figure}[!ht]
\begin{center}
\includegraphics[width=0.495\textwidth, clip ]{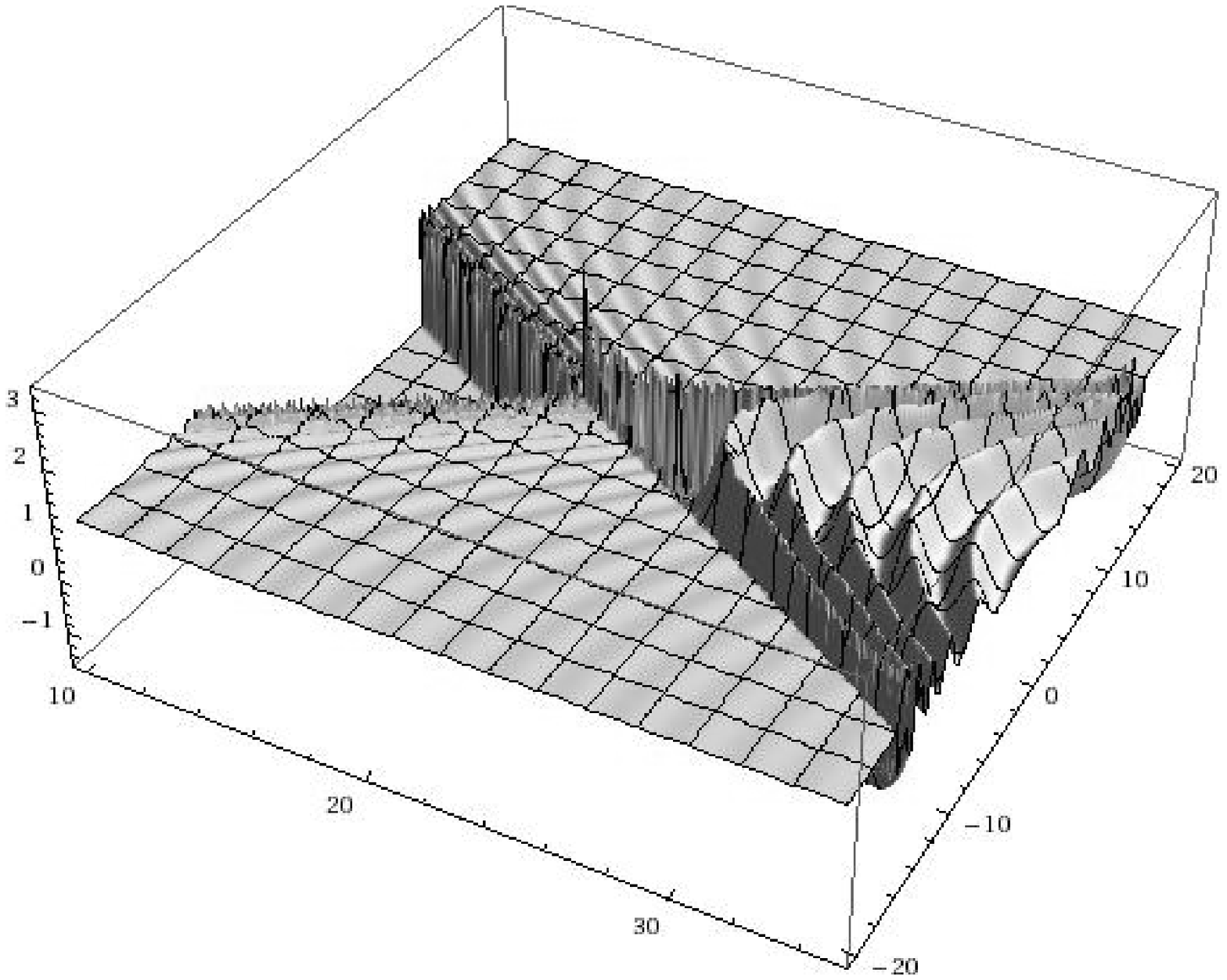}
\caption{
\label{fig:3D}
\small 3D plot corresponding to the bottom right plot of
Fig.~\ref{fig:coll}.  }
\end{center}
\end{figure}
This situation is demonstrated in Fig.~\ref{fig:coll_slice},
\ref{fig:coll} and \ref{fig:3D} where we display the results of a
simulation of the collision of two planar bubbles in a potential of
the form
\be
\label{eq:toy_pot}
V(\phi) = \lambda (\phi^2 - v^2)^2 + \eta \, v \phi (\phi^2- 3 v^2).
\ee
The scalar field obeys the two-dimensional equation of motion
\be
\partial_t^2 \phi - \partial_z^2 \phi= -\frac{dV}{d\phi},
\ee
and the initial conditions are chosen as bubbles with opposing
expansion directions with wall thickness $l_w$ and velocity $v_w$. For
a single bubble we use
\be
\phi(z,t_0) = \frac{\phi_c}2
\left[ \, 1 \pm \tanh \, \lp\gamma_w 
\left[ v_w (t-t_0) -(z - z_0) \right] / l_w \rp \, \right],
\ee
and we choose the wall thickness to be the value obtained in the thin
wall approximation ($l_w^{-1} = \sqrt{2 \lambda} v $). The parameter
$\eta$ quantifies the level of degeneracy of the two minima. We show
results for the two values $\eta=0.2$ and $\eta=0.6$. In the rather
symmetric case ($\eta=0.2$), the walls bounce back to the symmetric
phase. In the asymmetric case ($\eta=0.6$), the field stays in the
basin of attraction of the broken phase and starts oscillating around
it after a short while of slow roll behavior close to the
maximum. Notice that in both cases bubble walls are present after the
collision and store the predominant fraction of the vacuum energy. In
the asymmetric case, the reflected walls do not loose their energy
while expanding into the broken phase and they expand until they meet
another reflected wall and thermalize by scattering. We have checked
that we obtain similar results for a nearly conformal potential of type
(\ref{eq:potential}).  The 3D representation of the collision
corresponding to the bottom right plot of Fig.~\ref{fig:coll}
is shown in Fig.~\ref{fig:3D}.

In the bottom plots of Fig.~\ref{fig:coll}, we show the fraction of
the energy stored in gradient and kinetic energy of the scalar field,
since this is the relevant quantity determining the production of
winding configurations.  What we call ``kinetic" is the sum of the
gradient plus kinetic energy in the broken phase excluding the wall.
Note that the total energy in the asymmetric case is about a factor 3
larger than in the symmetric case (due to the larger difference in the
potential minima). Note also that most of the ``kinetic" energy in the
symmetric case actually results from the wall: When the bubble wall
changes direction, the wall becomes thicker and reaches into the region
that we attribute to the broken phase (see Fig.~\ref{fig:coll_slice}).
From these plots, it is clear that for nearly degenerate minima, there
is little energy transferred in kinetic energy of the scalar field
whereas for an asymmetric potential (c), a large fraction of the
energy ends up in gradient and kinetic energy of the Higgs.

We conclude that bubble collisions in an empty universe
as arising in a nearly conformal phase transition lead to a
situation that closely resembles the situation after low-scale hybrid
inflation: First, bubbles nucleate and expand. Then, the walls are
reflected and sweep space a second time. After the bubble wall has
passed a second time, the scalar vev is arranged close to the
symmetric phase but beyond the potential barrier of the asymmetric
potential (in the basin of attraction of the broken
phase). Subsequently the field approaches the broken minimum and
releases the potential energy while oscillating around it. After the
walls swept all space twice, the scalar field is everywhere in the
basin of attraction of the broken phase and reheating begins. The
bubble walls collided only once during this early stage and energy
drain into the fermionic sector is rather small.

In summary, the important point is that a large kinetic energy is
taken from the radion field and drawn into the Higgs sector  
(we are assuming that there is a sizable coupling between the Higgs and the radion fields). Like in
usual preheating the radion has time to oscillate before it decays.
This is put by hand in the hybrid inflation model of original cold
baryogenesis. In our setup, the radion plays the role of the
inflaton. However, we do not have to assume slow-roll since inflation
is due to supercooling.  Besides, another interesting feature is the
little hierarchy between the scale of conformal symmetry breaking and
the scale of EW symmetry breaking. Because the energy in the radion
sector is large relatively to the Higgs sector, it is easy to produce
Higgs winding number by pumping energy from the radion sector. 

\section{ Estimate of the baryon asymmetry \label{sec:baryo}}

 The production of Higgs-winding requires the occurrence of zeroes in the Higgs field. In the simulations of tachyonic preheating, this happens via the self-interactions of the Higgs field and the complexity of the dynamics in the $SU(2)$ gauge space. Our simplistic simulations do not include the gauge structure and we therefore cannot study the production of winding number at the end of the transition. The main point we made in the previous section is  that at the end of the transition, almost all the energy of the system
resides in the potential and kinetic energy of the scalar field. We believe this makes abundant Higgs winding production  very plausible once  the gauge structure is included.
Compared to the scenario of low-scale hybrid
inflation, there is significantly more energy in the scalar sector at
this stage since not only the Higgs but also the radion potential
energy fuels the scalar dynamics (in slow-roll inflation most of the
potential energy of the inflaton is inflated away).  In fact, the
difficulty might be not to have too much energy in the system and to
keep the reheat temperature low enough. The reheating temperature has
to be below the temperature $T_{EW}$ at which the electroweak symmetry
is restored and at which sphalerons are active in order to prevent the
washout of the baryon asymmetry.  In models of nearly conformal dynamics
at the TeV scale, whether the reheat temperature exceeds $T_{EW}$ 
actually depends on the Higgs and radion
masses, as discussed in
detail in \cite{Konstandin_Servant1}.

A second difference is in the specific initial conditions. In hybrid
inflation, these initial conditions are set by the spinodal
instability in the Higgs sector with random orientation in $SU(2)$
gauge space. The Higgs field approaches and oscillates in the Mexican
hat potential everywhere more or less at the same time but with a
different direction in $SU(2)$ space at different locations. This
large degree of disorder leads to the fact that thermalization in the
Higgs sector happens relatively fast~\cite{Felder:2000hj,
Skullerud:2003ki}.   In contrast, in our
simulations, the time when the Higgs approaches the potential minimum
depends on the position in relation to the colliding bubble walls, and
from our plots, the fields appear much more ordered. Clearly, this is
due to the fact that we neglect the $SU(2)$ gauge structure, while colliding bubbles
with different phases in gauge space lead to magnetic fields and a
certain amount of disorder~\cite{Kibble:1995aa}. Still, if two
colliding bubbles had the same orientation in gauge space, the problem
would effectively become one-dimensional. This should partially
suppress the early generation of Higgs winding. Furthermore, due to
the lower degree of disorder in gauge space, it is expected that the
thermalization of the scalar sector takes a longer time.

Having these differences in mind, we state a rough estimate of the
baryon asymmetry in cold electroweak baryogenesis as given in
\cite{Felder:2000hj, Tranberg:2006dg}. The estimate is based on the
fact that the CP violating operator in eq.~(\ref{eq:CPVop}) can be
interpreted (after partial integration) as a chemical potential of the
Chern-Simons number
\be
\int d^4x \,  \frac{1}{M^2} \phi^\dagger \phi\, \tilde F F \leftrightarrow
\int dt \, \mu_{cs} \, N_{cs},
\ee
of size
\be
\mu_{cs} \propto \frac{1}{M^2} \frac{d}{dt} \left< \phi^\dagger \phi\right>.
\ee
The resulting estimate for the baryon asymmetry reads
\cite{GarciaBellido:1999sv}
\be
\frac{n_B}{s} \propto 3 \times 10^{-5} \, \frac{v^2}{M^2} 
\left( \frac{T_{eff}}{T_{rh}} \right)^3,
\ee
where $T_{eff}$ is the effective temperature of the soft Higgs
modes 
\footnote{ Using the even more simplistic estimate of
\cite{Krauss:1999ng}, one arrives at a smaller asymmetry:
\be
\label{eq:eta_estimate}
\frac{n_B}{s} \propto 4 \times 10^{-6} \frac{v^2}{M^2} 
\left( \frac{v}{T_{rh}} \right)^3.
\ee
 }.
Ultimately, simulations in the context of inverted hybrid inflation
gave the result \cite{Tranberg:2006dg}
\be
\label{eq:eta_estimate2}
\frac{n_B}{s} \propto 3 \times 10^{-3} \, \frac{v^2}{M^2} ,
\ee
corresponding to an effective temperature of order $T_{eff}
\simeq 5 \, T_{rh}$.

This has to be confronted with the experimental constraint on the
operator (\ref{eq:CPVop}) that contributes e.g.~to the electric dipole
moment of the electron. The arising dipole moment has been estimated
to~\cite{Lue:1996pr}
\be
\frac{d_e}{e} \simeq \frac{m_e \sin^2 (\theta_W)}{8 \pi^2} 
\frac{1}{M^2} \log \frac{M^2 + m_H^2}{m_H^2}.
\ee
For a Higgs mass $m_H \sim 200$ GeV, the upper bound on the electric
dipole moment~\cite{Regan:2002ta}, $d_e < 1.6 \times 10^{-27} \, e \,
{\rm cm}$, leads to the constraint $M \gtrsim 14$ TeV. Comparison with
the estimates (\ref{eq:eta_estimate}) and (\ref{eq:eta_estimate2})
shows that in the present setup baryogenesis is possible as long as
the radion fuels the scalar sector with enough energy, which is not a
problem as demonstrated in the previous section.

The form of the new CP violation source will most probably not have a
large impact on the result as long as it solely involves the Higgs and
gauge fields, as the simulation with a different source in
Ref.~\cite{Tranberg:2009de} indicates. Nevertheless, one has to admit
that the above estimate can at best predict the early production of
Chern-Simons number while for the final baryon asymmetry also the evolution
of the Higgs winding plays an important role (this is nicely demonstrated in
\cite{Tranberg:2006dg}).  An ultimate judgment can at
this stage only come from simulations.

\section{Conclusion \label{sec:dis}}

In the last twelve years, the scenario of cold baryogenesis has been
based on a (not particularly motivated) hybrid inflation
model. Somehow, it had not been thought of in the context of reheating
from bubble collisions.  Indeed, first-order phase transitions have
been commonly studied based on standard polynomial potentials, in
which case the amount of supercooling is not sufficient for cold
baryogenesis to work.  However, nearly conformal potentials coupled to
the Higgs sector, as motivated by a dynamical solution to EW symmetry
breaking, can lead to the ideal conditions for a cold electroweak
phase transition \cite{Konstandin_Servant1}.

We have shown that during bubble collisions, there can be very
efficient energy transfer from the bubble wall to the classical Higgs
gradient and kinetic energy, hence allowing the production of a large
population of winding configurations and non-zero Chern-Simons number.
Like in usual EW baryogenesis, the source of baryon number violation
is purely Standard Model-like. In this sense, the scenario we propose
is based on rather conservative assumptions.  The mechanism relies on
classical dynamics of the gauge and Higgs field of the Standard Model
at the EW phase transition when assuming non-standard CP violating
effects (typically through a dimension-six operator involving the
time-variation of the Higgs field). Most of the baryon asymmetry is
produced when the Higgs is rapidly rolling down its potential, while
the gauge fields relax to cancel the gradient energy of the Higgs
field.

For the mechanism to work, it is crucial that it takes place in
a cold universe, where the coherent bosonic fields can evolve
unhindered by the thermal
plasma~\cite{GarciaBellido:1999sv}. Besides, the Universe has also to
be sufficiently cold after the EW phase transition so that thermal
sphalerons production does not wash out the baryon asymmetry produced
during preheating. This is the case provided that the radion is relatively light and/or that
the Higgs is heavy
enough, so that the reheat temperature is naturally below the
sphaleron freeze out temperature \cite{Konstandin_Servant1}.

 
Our arguments remain qualitative and we believe they motivate further
study. It would be interesting to make numerical simulations similarly
to what has been done for hybrid inflation~\cite{Rajantie:2000nj,
Felder:2000hj, GarciaBellido:2002aj, Smit:2002sn,
GarciaBellido:2003wd, Tranberg:2003gi}. At first sight, this
seems  impossible in the context of first-order phase
transitions, since the problem involves two widely separated scales:
The size of the bubbles and the electroweak scale. However, in the
case of a nearly conformal potential, the size of the bubbles is no longer
important after the first stage of  bubble collisions (see
Fig.~\ref{fig:coll}) and the simulation of Higgs winding production
seems feasible.

Finally, we stress that the underlying theoretical framework we have
in mind can be studied in an extra-dimensional context. Holography has
proved to be very useful to calculate quantities in strongly coupled
theories. We note that there may be interesting related physics
concerning the TeV mass skyrmions of \cite{Pomarol:2007kr} which may be
abundantly produced at preheating in the way we have shown and could
either play a role in baryogenesis or dark matter generation (as they
can be stable).

\section*{Acknowledgments}
This work is supported by the ERC starting grant Cosmo@LHC (204072).

\end{document}